\documentclass{article}

\usepackage[english]{babel}

\usepackage[a4paper,top=2cm,bottom=2cm,left=3cm,right=3cm,marginparwidth=1.75cm]{geometry}

\usepackage{amssymb}
\usepackage{siunitx}
\PassOptionsToPackage{hyphens}{url}\usepackage{hyperref}
\usepackage{cleveref}
\usepackage[utf8]{inputenc}
\usepackage{csquotes}
\usepackage{booktabs}
\usepackage{longtable}
\usepackage{adjustbox}
\usepackage{array}
\usepackage{url}
\usepackage{titlesec}
\usepackage{authblk}
\usepackage{xcolor} 

\titleformat{\subsection}
  {\mdseries\itshape\large} 
  {\thesubsection}{1em}{} 

\usepackage[english]{babel}
\usepackage[numbers,authoryear]{natbib}
\setcitestyle{round}

\let\textcite\citet
\let\parencite\citep

\crefformat{figure}{#2Figure~#1#3}
\Crefformat{figure}{#2Figure~#1#3}
\crefformat{table}{#2Table~#1#3}
\Crefformat{table}{#2Table~#1#3}
\crefformat{section}{#2Section~#1#3}
\Crefformat{section}{#2Section~#1#3}

\author[1,2]{Shah Rukh Qasim}
\author[2]{Patrick Owen}
\author[2]{Nicola Serra}

\affil[1]{shahrukh.qasim@physik.uzh.ch (corr. author)}
\affil[2]{University of Zurich, Physics Institute, Zurich, Switzerland}

\title{Physics Instrument Design with Reinforcement Learning}

\begin{document}
\maketitle

\begin{abstract}
We present a case for the use of Reinforcement Learning (RL) for the design of physics instrument as an alternative to gradient-based instrument-optimization methods. It's applicability is demonstrated using two empirical studies. One is longitudinal segmentation of calorimeters and the second is both transverse segmentation as well longitudinal placement of trackers in a spectrometer. Based on these experiments, we propose an alternative approach that offers unique advantages over differentiable programming and surrogate-based differentiable design optimization methods. First, Reinforcement Learning (RL) algorithms possess inherent exploratory capabilities, which help mitigate the risk of convergence to local optima. Second, this approach eliminates the necessity of constraining the design to a predefined detector model with fixed parameters. Instead, it allows for the flexible placement of a variable number of detector components and facilitates discrete decision-making. We then discuss the road map of how this idea can be extended into designing very complex instruments. The presented study sets the stage for a novel framework in physics instrument design, offering a scalable and efficient framework that can be pivotal for future projects such as the Future Circular Collider (FCC), where most optimized detectors are essential for exploring physics at unprecedented energy scales.
\end{abstract}

\section{Introduction}
\label{sec:introduction}
The study of the fundamental nature of the universe relies on very complex physics detectors and instruments in the modern era. These devices, essential for probing the smallest constituents of matter, detecting elusive particles, and testing fundamental theories, are often highly specialized and extraordinarily expensive. For instance, only the construction and materials for the Compact Muon Solenoid (CMS) experiment at the Large Hadron Collider (LHC) amounted to 500 million Swiss Francs. Designing these instruments is a monumental challenge, as every component must be optimized for maximum efficiency, sensitivity, and precision to ensure the greatest scientific output. The design parameter space of these instruments is highly complex and multi-dimensional, requiring specifications for a wide range of variables, including the type, composition, and geometry of the materials used, as well as their spatial arrangement. Additionally, factors such as thermal stability and radiation hardness must be considered, all while ensuring that the components can withstand the extreme operational conditions often encountered in high-energy physics experiments. This intricate interplay of parameters forms a vast and challenging design optimization problem, where sometimes small adjustments can have significant impacts on the overall performance and sensitivity of the instrument.

Recently, the use of Machine Learning (ML) has exploded in different areas for solving a vast set of optimization tasks, including high energy physics as well. Machine learning algorithms, particularly those involving deep learning, have proven highly effective in analyzing large datasets, identifying complex patterns, and making predictions. In high energy physics, where experiments often generate immense amounts of data, machine learning techniques are invaluable for tasks such as particle identification, event classification, and anomaly detection. These approaches not only enhance the precision of experimental results but also accelerate the process of discovery by automating time-consuming calculations and allowing researchers to explore new theories and phenomena with greater efficiency.

Building on the advances in the machine learning techniques, investigators recently began exploring machine learning also for the optimization of physics instruments. Here, the focus has been on fine tuning the design parameters along the direction of the gradient, similar to how neural networks are trained. This is either done through differentiable programming i.e., modifying the software code itself to make it differentiable or through a differentiable surrogate. In the later case, a neural network is trained as a surrogate for the complex simulator. By nature, the neural network is differentiable and one can backpropagate through it to tune the design parameter to increase the performance of the instrument.


Differentiable optimization methods, while useful, present two significant challenges for detector design and optimization. First, these methods tend to be inherently local, often leading to solutions trapped in local minima. This issue is particularly pronounced when tuning design parameters for instruments, where the parameter space is typically more structured and less random. In contrast, neural network training often deals with highly randomized parameters, where techniques like stochastic gradient descent can more effectively navigate out of local minima. Alternative approaches, such as Bayesian Optimization, mitigate the issue of local minima by exploring the parameter space more broadly. However, they tend to be computationally less efficient and don't scale to high-dimensional problems. Moreover, both differentiable and Bayesian Otimization methods are limited to tuning pre-defined parameters within an established model framework. For instance, while these methods are effective in adjusting the thickness of different layers in a calorimeter, they are unable to address the more complex, combinatorial nature of instrument design. In the case of calorimeters, neither approach allows for the optimization of the material selection or the number of the detector layers, which remains a crucial challenge in automated design of such complex systems.

The limitations of these methods become evident in the context of real-world detector design. As an example, back in the early 2000's, the design of the LHCb detector was reoptimized due to cost constraints~\parencite{reoptimized_lhcb}. The tracking stations at the dipole magnets were removed. This reduced multiple scattering and actually improved the performance of the detector (despite being cheaper). Similarly, at the NA62 experiment at CERN~\parencite{na62}, a secondary downstream calorimeter (HASC2) was installed as it provided further veto capabilities to reject the unwanted backgrounds~\parencite{na62_status}. This was discovered during the experiment's run, though it could, in principle, have also been identified in the simulation. Differentiable methods of optimized designs of physics experiments will not be able to find such solutions.

To overcome these challenges, this work instead explores the use of RL for instrument design, as RL's exploratory nature helps avoid local minima by balancing exploration and exploitation. Second and more importantly, unlike gradient-based methods, RL is not restricted to a predefined set of parameters and can design instruments from scratch while navigating complex, non-convex, and combinatorial spaces. This approach has already proven successful in other fields, such as chip floor planning~\parencite{mirhoseini2021graph}. Here, the authors used Proximal Policy Optimization~\parencite{ppo_cite} to place various chip elements one after the other. In under six hours, the RL method outperforms the human design, which takes weeks manually. The method is now actively used at Google for the design of their TPUs (Tensor Processing Units). The methodology has also been applied on other problems such as material design ~\parencite{guo2021artificial}, and for molecule design~\parencite{zhou2019optimization}.

Through the empirical studies as well as the arguments presented, the aim is to demonstrate the feasibility of using RL for the automated design of physics experiments, and identify the most effective RL strategies for different types of physics problems and  This will set the stage for the development of new physics instruments by using machine learning to explore and optimize complex design configurations from scratch. This will be instrumental for future projects like the Future Circular Collider (FCC), where machine learning based instrument design will prove highly invaluable to allow probing new physics at unprecedented energy scales.

\section{Related work}
The most notable use case in physics instrument optimization has been the optimization of the Muon Shield for the SHiP Experiment. Using a series of tapered magnets, the goal of Muon Shield is to reduce the muon flux by six orders of magnitude. Each of the magnets is parameterized by seven values, which represent the dimensions (such as length, thickness, etc). The problem is non-trivial and formulates a complex optimization task. A high energy muon can be deflected in a longer yet simpler curved path but a lower-energy muon might go in a zigzag trajectory needing multiple magnets. The initial design iterations were performed with the help of Bayesian Optimization~\parencite{baranov2017optimising}. Later, \textcite{shirobokov2020black} proposed a method to employ neural surrogates to allow gradients on the design parameters. Here, a Generative Adversarial Network was trained as a surrogate model for the Geant4 simulator~\parencite{geant4_cite}. The surrogate is a neural network, hence one can back-propagate through it fully to tune the design parameters. 

Another example of the use of machine learning for detector optimization has been at the proposed Electron Ion Collider (EIC)~\parencite{accardi2016electron}. Here, dual ring-imaging Cherenkov (RICH) detector was optimized for improved $\pi$/$K$ separation. This was also performed via Bayesian Optimization. Later, detector optimization for one of the detectors at the EIC called EIC Comprehensive Chromodynamics Experiment (ECCE) was also performed via Bayesian Optimization.
\textcite{dorigo2023toward} provides a comprehensive overview of current and past efforts in applying machine learning for instrument optimization, with a primary focus on differentiable methods (but also discusses Bayesian Optimization). It is a remarkable read and delves deeply into the advocated methodology as well as many associated examples. This publication originates from the \textcite{mode_collaboration}, a group of physicists and machine learning practitioners dedicated to advancing design optimization through differentiable programming. The collaboration's objectives encompass a broad range of applications, from optimizing detectors in particle physics to industrial uses of radiation detection technology.

The first example of differentiable programming for detector optimization is presented by \textcite{Strong2024}. The authors presented an alternate Software to Geant4 with gradient implementations of the various functions used for the muon topography. Similarly, \textcite{jans_differentiable_geant4} explored whether operator overloading can be used to make Geant4 directly differentiable.

\section{Reinforcement Learning}
\label{sec:method}
\begin{figure}[ht!]
    \centering
    \includegraphics[width=0.9\textwidth]{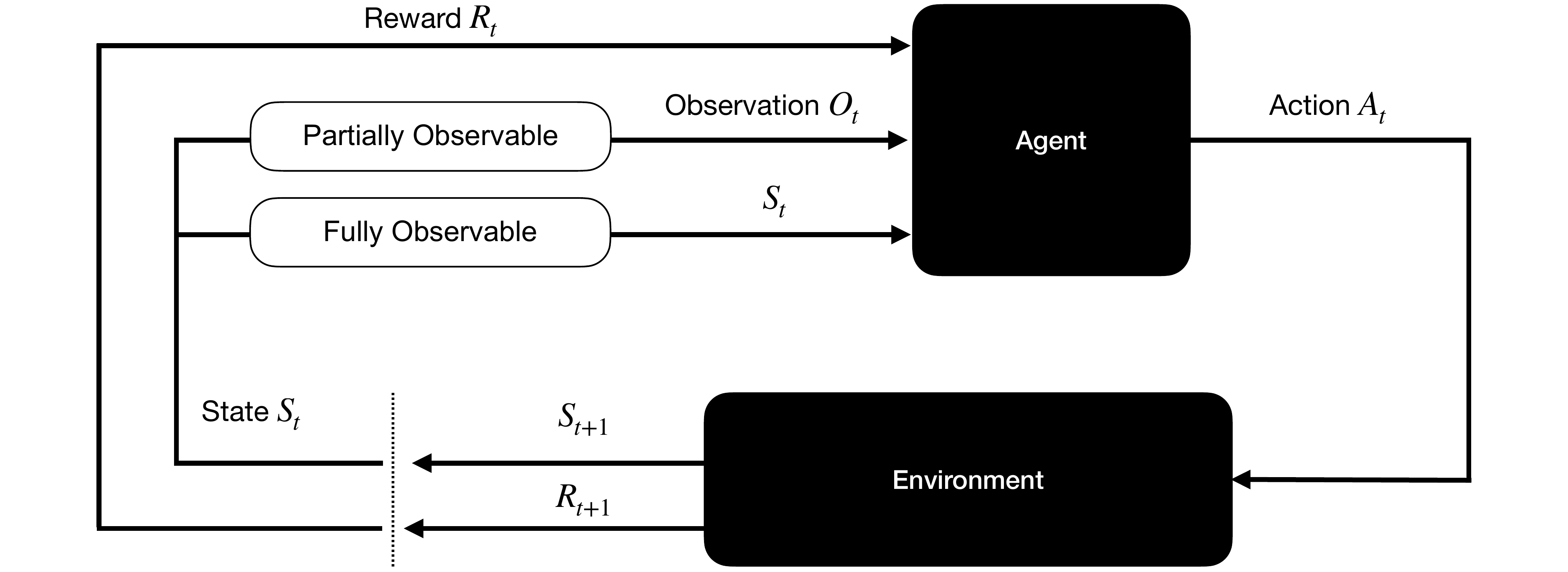}

    \caption{Agent-environment interaction in Reinforcement Learning (adapted from \textcite{sutton2018reinforcement})}

    \label{fig:what_is_rl}
\end{figure}

Reinforcement Learning (RL) is a branch of machine learning focused on training agents to make sequences of decisions by interacting with an environment (illustrated in Figure~\ref{fig:what_is_rl}). The process unfolds as a series of interactions between the agent and the environment, described mathematically as follows:

\begin{itemize}
    \item \textbf{State ($S_t$)}: At each time step $t$, the agent observes the state of the environment, denoted as $S_t$. The state encapsulates all the relevant information needed for decision-making at that point in time.  However, this state may not always be fully accessible to the agent.
    \item \textbf{Observation ($O_t$)}: Instead of directly observing the full state $S_t$, the agent may receive an observation $O_t$, which provides partial information about the environment. This is common in scenarios where the environment is \textit{partially observable}, meaning the agent has limited or noisy access to the underlying state. In contrast, in a \textit{fully observable} environment, the observation $O_t$ is equivalent to the true state $S_t$. \item \textbf{Action ($A_t$)}: Based on the observed information (either $S_t$ in fully observable settings or $O_t$ in partially observable settings), the agent selects an action $A_t$ from a set of possible actions. The choice of action is determined by the agent's policy, which is the strategy it follows.
    \item \textbf{Reward ($R_t$)}: After executing action $A_t$, the agent receives feedback in the form of a reward $R_t$. This scalar value indicates how good or bad the action's outcome was in relation to the agent's objective.
    \item \textbf{Next State ($S_{t+1}$)}: As a result of the action $A_t$, the environment transitions to a new state $S_{t+1}$, which the agent observes at the next time step.
\end{itemize}

This interaction sequence can be written as:
\[
(S_t, A_t, R_t, S_{t+1})
\]
The agent's goal is to maximize the cumulative reward it collects over time, often referred to as the \textit{return}. The return \(G_t\) is defined as the total discounted reward from time step \(t\) onward:
\begin{equation}
\label{eqn:rl_return}
G_t = R_{t+1} + \gamma R_{t+2} + \gamma^2 R_{t+3} + \cdots = \sum_{k=0}^\infty \gamma^k R_{t+k+1}~\text{,}
\end{equation}

where \(\gamma \in [0, 1]\) is the discount factor that determines the present value of future rewards. A smaller \(\gamma\) places less emphasis on future rewards, while \(\gamma = 1\) corresponds to no discounting.

The agent selects actions based on a policy \(\pi\), which is a mapping from states or observations to probabilities of actions. In a fully observable environment, the policy depends directly on the true state \( S_t \), and is defined as:
\[
\pi(a|s) = P(A_t = a | S_t = s),
\]
where \(\pi(a|s)\) denotes the probability of taking action \(a\) in state \(s\). However, in partially observable environments, the policy is based on the agent's observation \( O_t \) or its belief about the state.

The optimal policy \(\pi^*\) is the policy that maximizes the expected return from each state or observation. For fully observable environments, it is expressed as:
\[
\pi^*(s) = \arg\max_\pi \mathbb{E}_\pi \left[ G_t | S_t = s \right],
\]
while in partially observable settings, the expectation may be conditioned on the observation \( O_t \) or the agent's belief about \( S_t \).

The reward \(R_t\) is a scalar signal provided by the environment as feedback for the action \(A_t\). It represents the immediate benefit or cost associated with the action and guides the agent toward its objective. In partially observable environments, the agent must infer the underlying state \( S_t \) from observations \( O_t \) to learn an effective policy.

Together, the components \((O_t, S_t, A_t, R_t, O_{t+1}, S_{t+1})\) define the dynamics of the agent's interaction with the environment. In fully observable scenarios, these dynamics are modeled as a Markov Decision Process (MDP). In partially observable scenarios, they are generalized to a Partially Observable Markov Decision Process (POMDP), where the agent maintains a belief state—a probabilistic representation of the true state based on its observations and past actions.

In RL, the learning paradigm relies heavily on trial and error, with agents actively interacting with their environment to refine their decision-making strategies. Unlike traditional learning methods, where learning is often passive, RL agents must balance exploiting known strategies that yield high rewards and exploring new actions that might lead to better long-term outcomes. This balance, known as the exploration-exploitation trade-off, is a cornerstone of RL and is often implemented through strategies like $\epsilon$-greedy policies or entropy-based exploration. RL's inherent exploratory nature helps agents avoid getting trapped in local minima, a common issue in optimization techniques that focus on refining parameters locally. By sampling a wide range of states and actions, RL agents can escape suboptimal regions and discover global optima. Moreover, RL does not assume a fixed structure for the optimization landscape, enabling exploration across vast and complex spaces, including non-convex, combinatorial, and highly discontinuous environments. This adaptability allows RL to tackle challenges in dynamic and intricate settings, making it a powerful framework for solving complex decision-making problems.

To solve these problems effectively, RL algorithms use a variety of techniques, such as value-based methods (e.g., Q-learning), policy-based methods, and hybrid approaches like actor-critic algorithms. Value-based methods focus on estimating the value of being in a particular state, helping the agent understand the long-term benefits of certain actions. Policy-based methods, on the other hand, focus on directly learning a policy for action selection, which can be beneficial in environments with continuous action spaces. Combining these approaches, actor-critic algorithms enable an agent to simultaneously estimate the value of states and improve its policy, offering a more robust way of handling complex decision-making scenarios.

Beyond theoretical appeal, RL has demonstrated remarkable success in real-world applications, from robotics, where agents learn motor skills or navigation tasks autonomously, to video games and simulations, where RL agents can achieve superhuman performance. Moreover, RL has been applied in resource management, financial trading, and even healthcare, where personalized treatment strategies are developed based on patient data. Notably, RL is integral to systems like AlphaGo~\parencite{alpha_go} and AlphaZero~\parencite{alpha_zero}, where agents learn to master games like Go and chess without human input, relying purely on self-play.

RL is particularly effective for problems where the solution space is intricate or difficult to model with gradient-based methods, such as instrument design, where multiple design configurations need to be explored simultaneously. This method has already been employed for other problems (in particular, chip placement).

\section{Empirical Studies}
\begin{figure}[ht!]
    \centering
    \includegraphics[width=1.0\textwidth]{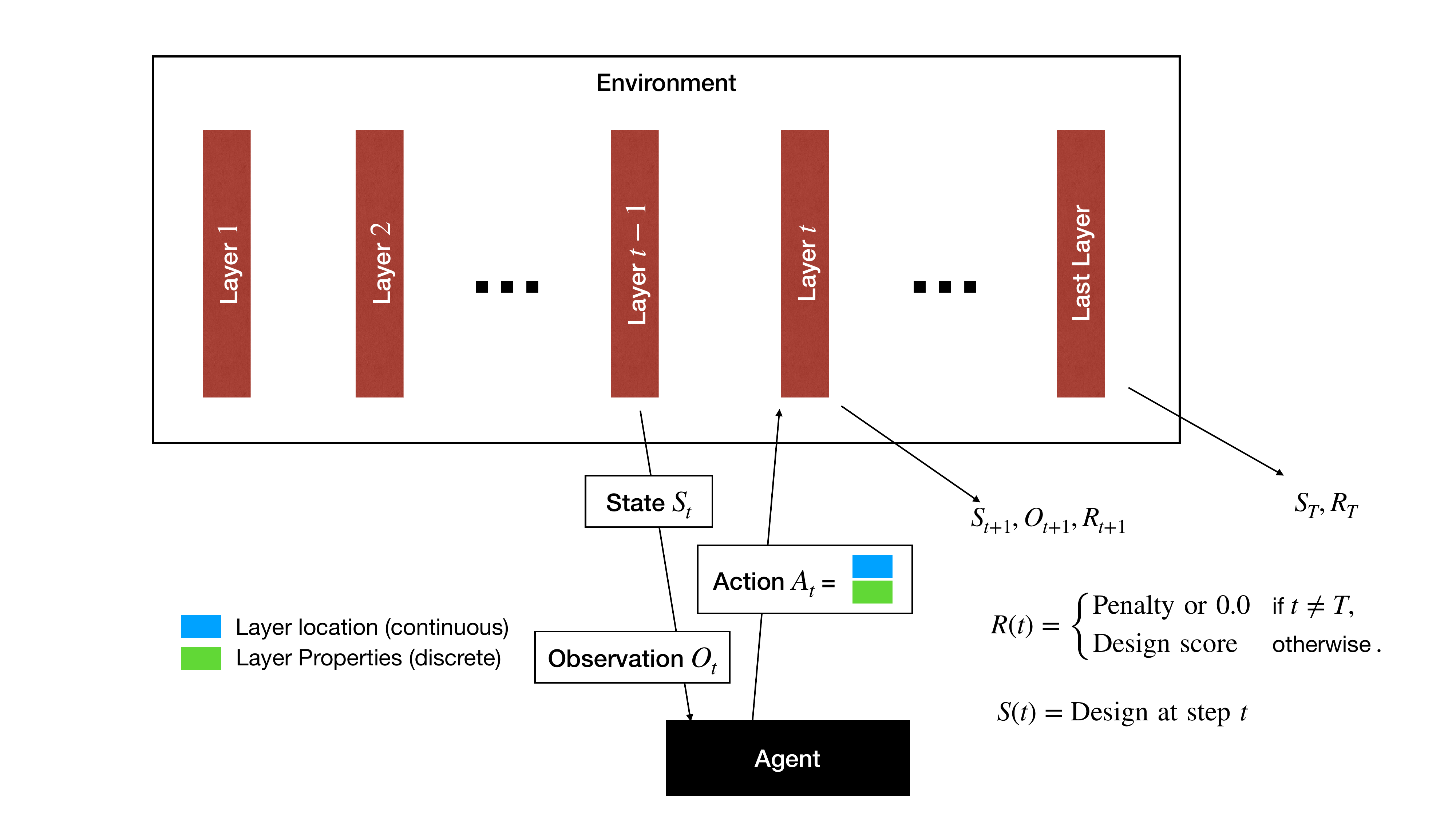}

    \caption{Reinforcement Learning for instrument design}

    \label{fig:rl_for_instrument_design}
\end{figure}

Two instrument design studies have been conducted to demonstrate the applicability of the proposed methodology. The first is longitudinal segmentation of calorimeters, which is presented in Section~\ref{sec:calo_design_study}. The second is longitudinal segmentation as well as transverse segmentation of tracking stations in spectrometers and is discussed in Section~\ref{sec:spectrometer_design_study}. The results are also discussed in the same sections. In both experiments, the agent is tasked with placing the layers one after the other until either the design is complete or the agent runs out of resources. This is demonstrated in Figure~\ref{fig:rl_for_instrument_design}. The continuous action determines $\Delta z$ until the next layer of the instrument and a discrete action chooses the parameters of the layer, such as thickness or granularity. At the final time step, the design score ($S$) of the instrument is computed and in the intermediate steps, the awarded reward is $0.0$ except in cases where the layers placed are too close to each other, in which case a penalty is awarded. Full episode Monte Carlo is used for optimization as opposed to Temporal Difference Learning to simplify the training process, and $\gamma$ (Equation~\ref{eqn:rl_return}) is taken as $0.0$.

In both experiments, we utilize Proximal Policy Optimization (PPO)~\parencite{ppo_cite}, an RL algorithm designed for efficient and robust policy optimization. PPO is an on-policy algorithm, meaning it learns exclusively from the most recent interactions with the environment rather than employing experience replay. A defining feature of PPO is its clipped objective function, which prevents excessively large updates to the policy, ensuring stable and reliable learning while allowing for incremental improvements. This simplicity and stability make PPO well-suited for tasks requiring flexible action-space handling, such as instrument design. Moreover, PPO supports both discrete and continuous action spaces, further emphasizing its versatility. In our implementation, the policy and value networks are two independent fully connected neural networks, each with two $U$-dimensional hidden layers, ensuring a clean separation of learning objectives without weight sharing. $U$ is 256 in the calorimeter design and $512$ in the spectrometer design.

\subsection{Calorimeter Design}
\label{sec:calo_design_study}
In particle physics experiments, a calorimeter is one of the main methods used to measure the energy of particles, with the other being a spectrometer. A calorimeter measures energy by fully absorbing the particles, whereas a spectrometer determines momentum by analyzing the deflection of particles in a magnetic field, based on their momentum and charge. Calorimeters provide a direct measurement of total energy, while spectrometers are used to infer momentum through particle trajectories.

The calorimeters are generally built as sampling devices where active readout layers and absorber materials are alternated. Generally the materials used are also different. However, in this study, a homogeneous sampling calorimeter was used. The simulations were performed using Geant4 and all the resulting data was captured, including the individual deposits as well as their spatial locations. Given that it is a sampling calorimeter, it allowed reuse of the same set of simulations across designs, saving computation time. Generally resampling is also an expensive operation, however, CUDA was leveraged to deploy the simulation on a Nvidia GPU, enabling testing of one design in $\sim 0.02$~seconds. When testing any configuration, first calibration needs to be performed, which was done over 50000 samples consisting half of photons and half of charged pions using similar strategy as what was used by ~\textcite{qasim2022end}.

The number of cumulative budget of the active layers is capped at $6.0$, with the RL agent having control over their placement, deciding where to position each active layer sequentially. Further, the agent can also choose the type of sensor from a set $\{Type 1, Type 2, Type 3\}$, where type 1 the cheapest and least performant and type 3 is the most performant but most expensive. The three types have a budget of $0.12$, $0.15$ and $0.20$ respectively. Putting together three actions for the logits of the discrete action and one for the continuous action, the action space is four dimensional. The RL agent is then tasked with selecting the optimal combination of active layers by using a reward function. The current configuration of active layers that the agent has selected so far is the state of the environment. However, the experiment is designed to be partially observable. Therefore, only the longitudinal location the agent is currently at and the thickness budget used so far is employed as $O_t$. This implies that the observation space is a 2D vector.

The performance of the energy measurement in a calorimeter is quantified by mean-corrected energy resolution, defined as standard deviation of the response ($E_{\mathrm{pred}}/E_{\mathrm{true}}$) divided by the mean. A lower value signifies better performance. The score ($S$) is based on the energy resolution of the calorimeter for both hadronic and electromagnetic particles and is defined as

\begin{equation}
S\cdot10 = -\mathrm{max}(0, \Sigma_{\mathrm{em50}} - 8) - \mathrm{max}(0, \Sigma_{\mathrm{em100}} - 5) - \mathrm{max}(0, \Sigma_{\mathrm{had50}} - 25) - \mathrm{max}(0, \Sigma_{\mathrm{had100}} - 18)~\text{,}
\end{equation}

with $\Sigma$ representing the mean-corrected resolution for different particles in percentage. The resolution is capped at a maximum of $150\%$. The subscripts $\mathrm{em}$ and $\mathrm{had}$ represent the electromagnetic and hadronic particles, respectively and the subsequent number denotes the energy of the test particles in GeV. For measuring these metrics, $2500$ particles of each type are employed ($10^4$ in total, not including the calibration set). The reward function is linear in resolution and drops to zero once the design reaches the reference resolution, and this was chosen as value that is slightly better than what was roughly expected in a reasonable design. The design score is given as the reward for the agent only at the last time step. In the intermediate time steps, a penalty of $-0.5$ is awarded if the agent places two active layers less than $10$~mm apart. The action is also capped at $10$~mm. 

\begin{figure}[ht!]
    \centering
    \includegraphics[width=0.95\textwidth]{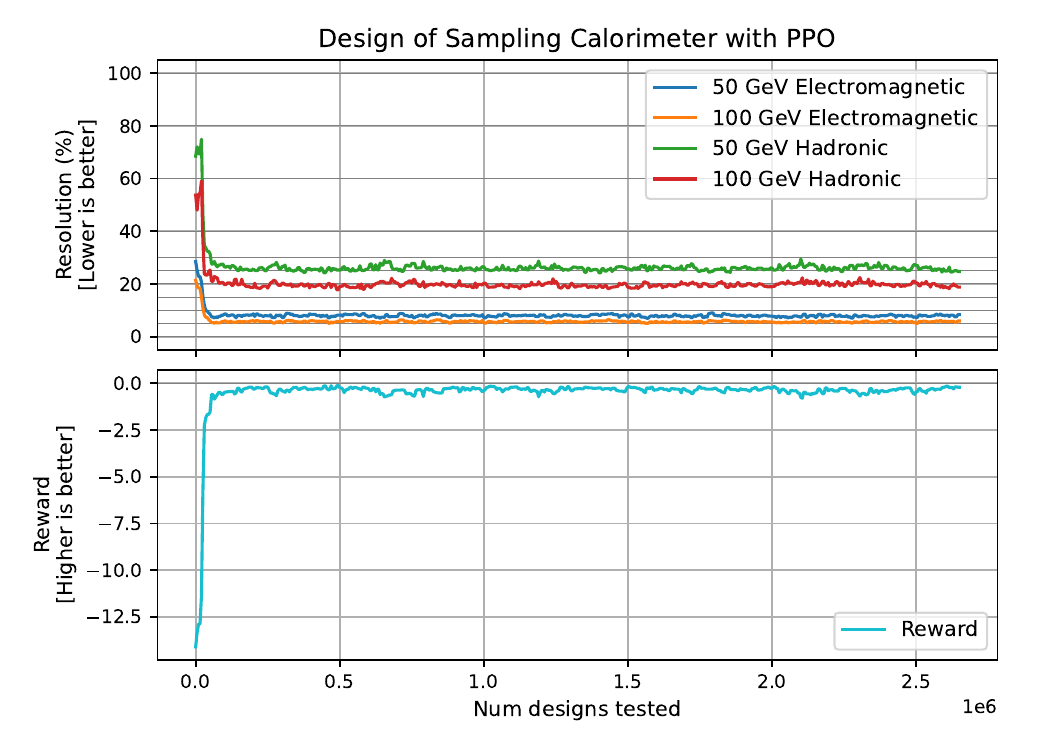}
    

    \includegraphics[width=0.95\textwidth]{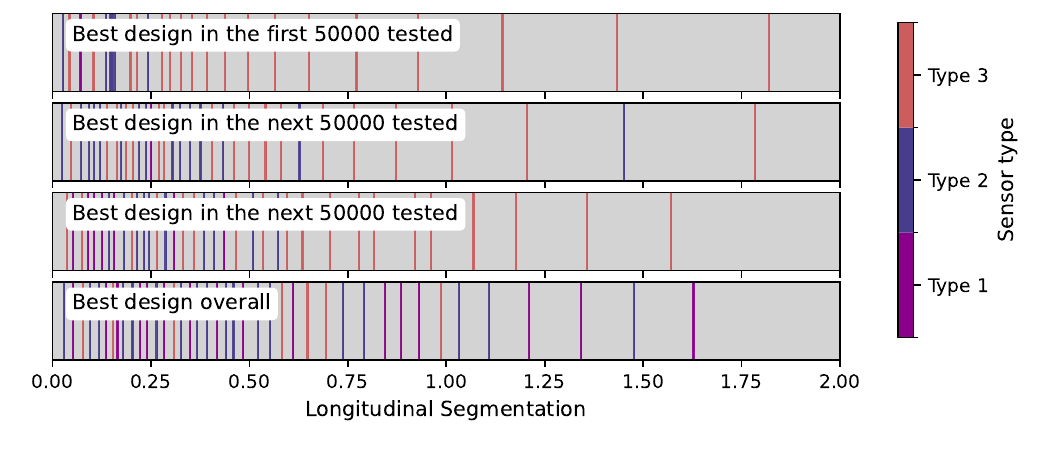}
    
    \caption{Design of uniform sampling calorimeter with proximal policy optimization (PPO). The top two plot shows the performance of the calorimeter as a function of iterations. The performance is represented by the resolution, calculated as $\frac{\sigma(E_{\mathrm{pred}}/E_{\mathrm{true}})}{\mu(E_{\mathrm{pred}}/E_{\mathrm{true}})}$, where $\sigma$ denotes the standard deviation and $\mu$ denotes the mean. The middle plot shows the cumulative reward during the episode (one design). In both of these figures, the the best design (as per the cumulative reward value) is chosen over intervals of 5000 designs and the resolution for different types of particles is plotted. The x-axis is shared between the two figures. The bottom plot shows the best designs found during different intervals}

    \label{fig:results_calo}
\end{figure}

Figure~\ref{fig:results_calo} visually demonstrates the learning process of the agent. The reward is shown as a function of the episodes (same as number of tested designs). It can be observed that the agent very quickly learns how to design good calorimeters in less than $100$k iterations. The resolution for the four test particles is also shown and it is negatively correlated with the reward function.

Figure~\ref{fig:results_calo} also illustrates four designs at different stages of training. Initially, the agent does not utilize the full sensor thickness budget, indicating a lack of understanding of the design requirements. Remarkably, without any prior information, the agent eventually learns the principles of calorimeter design, demonstrating the importance of placing more layers towards the front of the device. This reflects the distinct roles of the hadronic and electromagnetic sections of the device. Hadronic particles (e.g., protons, neutrons, and pions) primarily interact via the strong nuclear force, while electromagnetic particles (e.g., electrons and photons) interact through the electromagnetic force. The energy deposit profiles of these particles differ significantly, and a well-designed calorimeter must account for both\footnote{A common solution to this is to have two calorimeters, an electromagnetic calorimeter and hadronic calorimeter. We do not consider such a possibility in this study but is perfectly feasible within the framework of RL.}. Electromagnetic showers have a much shorter longitudinal length compared to hadronic showers, necessitating closer sensor spacing in the initial sections of the device. We refer to \textcite{Fabjan2003} for further discussion on the rich physics in calorimeters. Additionally, it is notable that the agent chooses layers with varying thicknesses to optimize the layout, allowing more layers to be placed and further demonstrating its ability to learn effective design strategies.

Table~\ref{tab:calo_results} quantitatively shows the performance metrics. The results are also compared to a baseline design. In the baseline design, the calorimeter is segmented at $\sim 22\cdot X_0$ of the material between the electromagnetic and hadronic sections, where $X_0$ refers to the radiation length. Afterwards, equal numbers of hadronic and electromagnetic layers are placed with a thickness of $200~\mu m$. The electromagnetic resolution is within the statistical error however, the hadronic resolution is significantly better in the calorimeter designed with RL. This result is expected as the nature of the physics of the hadronic processes is significantly more complex~\parencite{Fabjan2003} and requires more intelligent placement of the layers to get the most benefit.

\begin{table}[h!]
\centering
\begin{tabular}{|l|l|l|l|l|}
\hline
\textbf{} & \textbf{50 GeV EM} & \textbf{100 GeV EM} & \textbf{50 GeV Had} & \textbf{100 GeV Had} \\ \hline
\textbf{Baseline design} & $8.24 \pm 0.16$  & $5.94 \pm 0.12$  & $34.13 \pm 0.68$  & $24.48 \pm 0.49$  \\ \hline
\textbf{RL design} & $8.15 \pm 0.16$  & $5.83 \pm 0.12$  & $25.27 \pm 0.51$  & $17.79 \pm 0.36$  \\ \hline
\end{tabular}
\caption{Calorimeter resolution of the RL based design vs a reference baseline design. All numbers are in percentage}
\label{tab:calo_results}
\end{table}

\subsection{Spectrometer Design}
\label{sec:spectrometer_design_study}
A spectrometer is a instrument in particle physics experiments, designed to measure the momentum of charged particles by analyzing their trajectories as they traverse a magnetic field. Unlike calorimeters, which absorb particles to measure their total energy, spectrometers provide indirect energy measurements by leveraging the relationship between a particle's charge, momentum, and the curvature of its path in a magnetic environment. This curvature depends on the particle's momentum, with more energetic particles following less curved trajectories. Spectrometers are typically equipped with tracking stations, which are high-precision detectors placed at intervals along the particle's path. These stations record the position and trajectory of the particle before, during, or after its movement through the magnetic field, enabling the reconstruction of its momentum. Tracking stations often use advanced technologies, such as silicon strip detectors or drift chambers, to achieve the high spatial resolution necessary for accurate measurements. In this case we assume tracking stations that are made of silicon pixels, which are most useful not only for momentum resolution but also for pattern recognition. By combining data from multiple tracking stations, the spectrometer allows physicists to infer key properties of the particle, such as its momentum and charge.

\begin{figure}[ht!]
    \centering
    \includegraphics[width=0.95\textwidth]{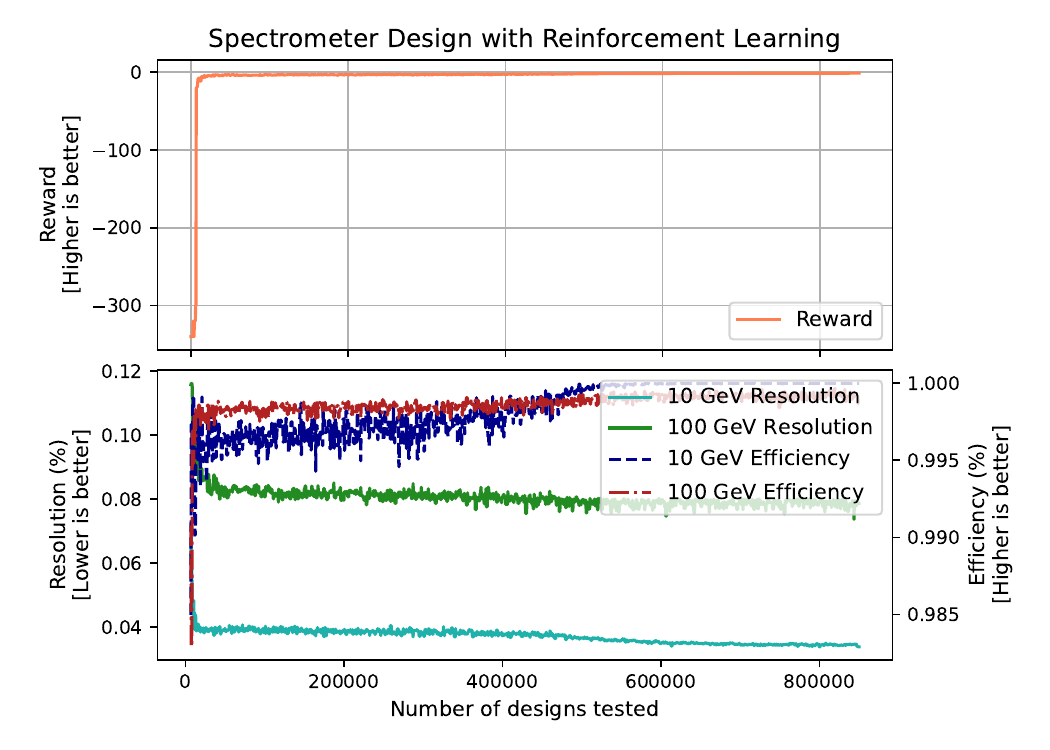}
    

    \includegraphics[width=0.95\textwidth]{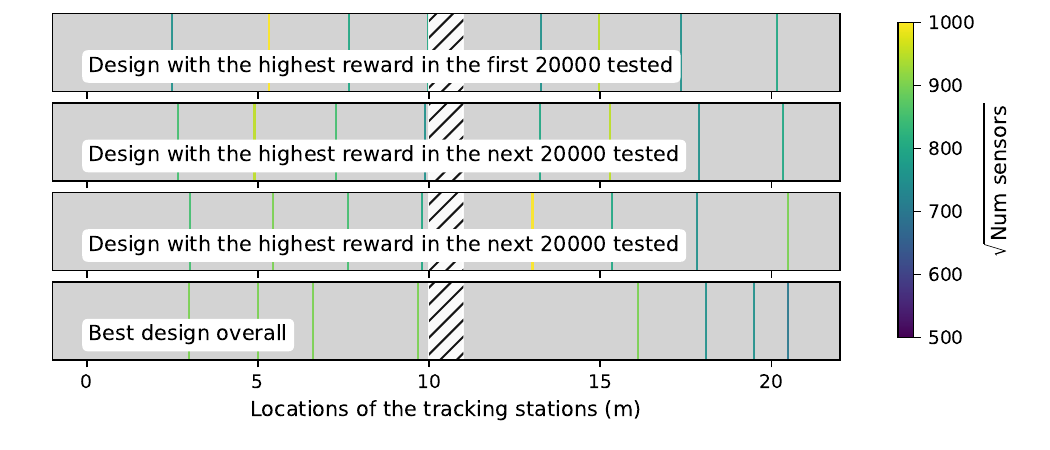}
    
    \caption{Design of spectrometer with Reinforcement Learning. The top two plot shows the performance of the calorimeter as a function of iterations. The performance is represented by the resolution, calculated as $\frac{\sigma(p_{\mathrm{pred}}/p_{\mathrm{true}})}{\mu(p_{\mathrm{pred}}/p_{\mathrm{true}})}$, where $\sigma$ denotes the standard deviation and $\mu$ denotes the mean. The top plot shows the cumulative reward during the episode (one design). In both of these figures, the the best design (as per the cumulative reward value) is chosen over intervals of 400 designs and the resolution for different types of particles is plotted. The x-axis is shared between the two figures. The bottom illustration shows the best designs found during different intervals during the training process. The magnet section is shown as the stripped section at the very middle and the number of sensors is shown as a color bar on the right side}

    \label{fig:results_spectro}
\end{figure}

In our experiment, a magnet is positioned at the center of the setup, spanning the region from $z = 10 \, \text{m}$ to $z = 11 \, \text{m}$ along the $z$-axis. The magnetic field is oriented along the $y$-axis with a uniform strength of $1 \, \text{T}$. A particle initially travels in a straight-line trajectory\footnote{Approximately straight line, considering the multiple scattering\label{footnote} discussed later} (called Region A) until it enters the magnetized region. Within this region, the particle experiences a Lorentz force acting along the $x$- and $y$-axes. Upon exiting the magnetic field, the particle resumes a straight-line trajectory in Region C\footnotemark[\value{footnote}]. The angle between the trajectories before and after the magnet can be approximated as
\begin{equation}
\label{eqn:sin_eqn}
p_{XZ} = \frac{0.3}{2 \sin\left(\frac{\theta}{2}\right)}~\text{.}
\end{equation}

Further, in the simulation, multiple scattering is also taken into account. The standard deviation of the multiple scattering is estimated as
\begin{equation}
\sigma_{\theta} = \frac{13.6~\text{GeV}}{\cdot p} \cdot Z \cdot \sqrt{\frac{x}{X_0}} \cdot 
\left(1 + 0.038 \cdot \ln\left(\frac{x}{X_0}\right)\right)~\text{,}
\end{equation}
with $Z$ being the atomic number of the material, $X_0$, the radiation length and $x$, the thickness~\parencite{Groom2000}. The numbers correspond to $200~\mu$m silicon sensors.

The agent is tasked with choosing the spatial locations where to place the tracking stations as well as granularity of each tracking station. The spatial location is chosen as the continuous action whilst the granularity selection task is modeled as a discrete action. The discrete action ($A^{\text{gran}}$) can be between $0$ and $9$ and the granularity of the placed sensor is then chosen as $(A^{\text{gran}} + 10)\cdot 500 A^{\text{gran}} + 10)\cdot 500)$ for $(x\times y)$. This means that each tracking station must have a minimum granularity of $500\times500$ pixels and a maximum of $1000\times 1000$. This makes the action space 12 dimensional. The resource budget is defined as having maximum $6$M sensors, at which point the episode terminates. The episode can also terminate if current $z$ location goes beyond the defined space i.e. $> 21.0~$m. The current configuration of the trackers that the agent has selected so far is the state of the environment. However, similar to the calorimeter design, the experiment is designed to be partially observable. Therefore, only the longitudinal location the agent is currently at and the number of layers placed so far is employed as $O_t$.

To allow momenta reconstruction, at least three tracking stations must be placed both in Region A and Region C. If this does not happen, the agent receives a $-400+20*N_{\text{correct}}$ reward, where $N_{\text{correct}}$ refers to the number of correctly placed stations, which maxes out at six (max of three in the two regions). Once the training reaches the stage where the agent knows it needs at least three tracking stations in the two regions, the design score is computed as follows 

\begin{equation}
S_{10} = -3 \cdot \big(95 - \min(\text{eff}_{10}, 95) + \max(\text{res}_{10}, 3) - 3\big)~\text{,}
\end{equation}

\begin{equation}
S_{100} = -3 \cdot \big(95 - \min(\text{eff}_{100}, 95) + \max(\text{res}_{100}, 8) - 8\big)~\text{,}
\end{equation}

\begin{equation}
    S = S_{10} + S_{100}~\text{.}
\end{equation}

Here, eff refers to efficiency which is defined as the fraction of the test particles where $0.5 < p_{\text{measured}}/p_{\text{true}} < 2.0$. This would occur in case of a mismatch in the pattern recognition (described below). The score is designed so the algorithm focuses on the resolution and efficiency is only taken into account if it is sufficiently large and hence, the associated penalty with efficiency drops to $0$ if it is higher than $95\%$. The design score $S$ is then given as the reward at the final time step, which also includes the penalty for missing tracking stations. In the intermediate time steps, a penalty of $-50$ is awarded as a reward if two tracking stations are placed too close to each other (i.e. $<0.1$~m). The test particle always has all the momenta in the direction of $z$ while $10$ background particles are also placed around uniformly distributed in the azimuthal range of $\left(0, \text{arctan}\left(0.1/20.0\right)\right)$. The distribution of the background particle is symmetric in the polar angle.

\begin{figure}[ht!]
    \centering
    \includegraphics[width=0.45\textwidth]{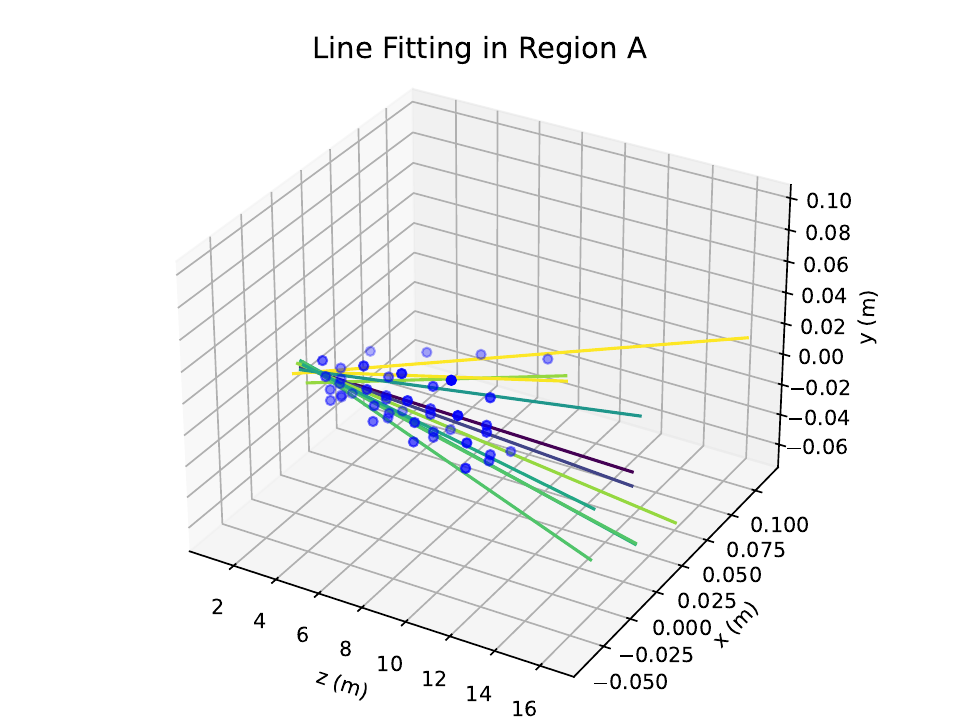}
    \includegraphics[width=0.45\textwidth]{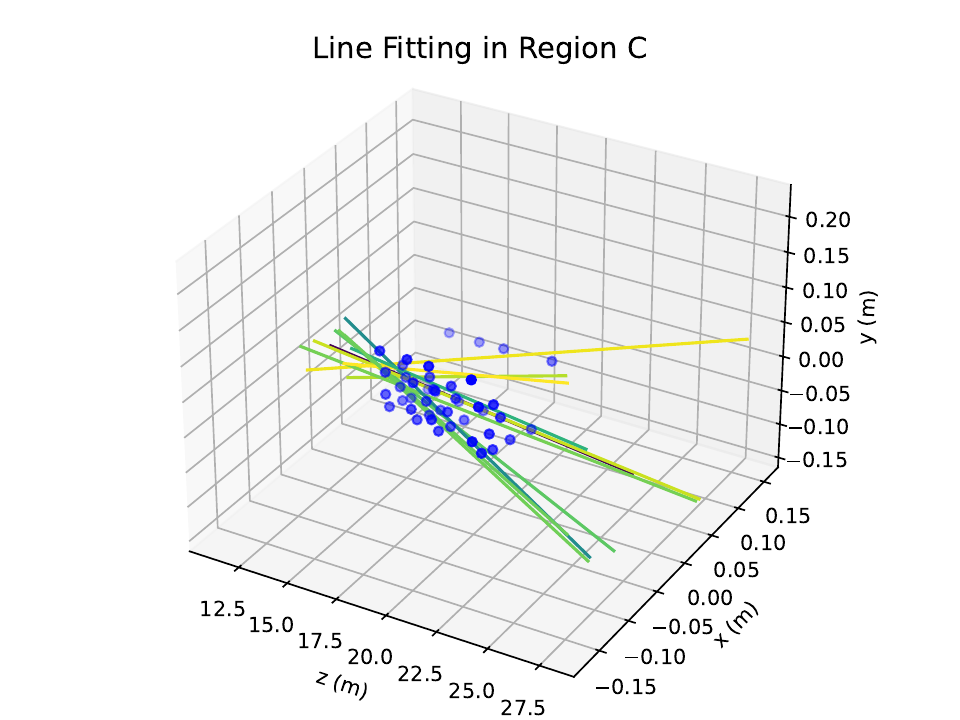}\\
    \includegraphics[width=0.45\textwidth]{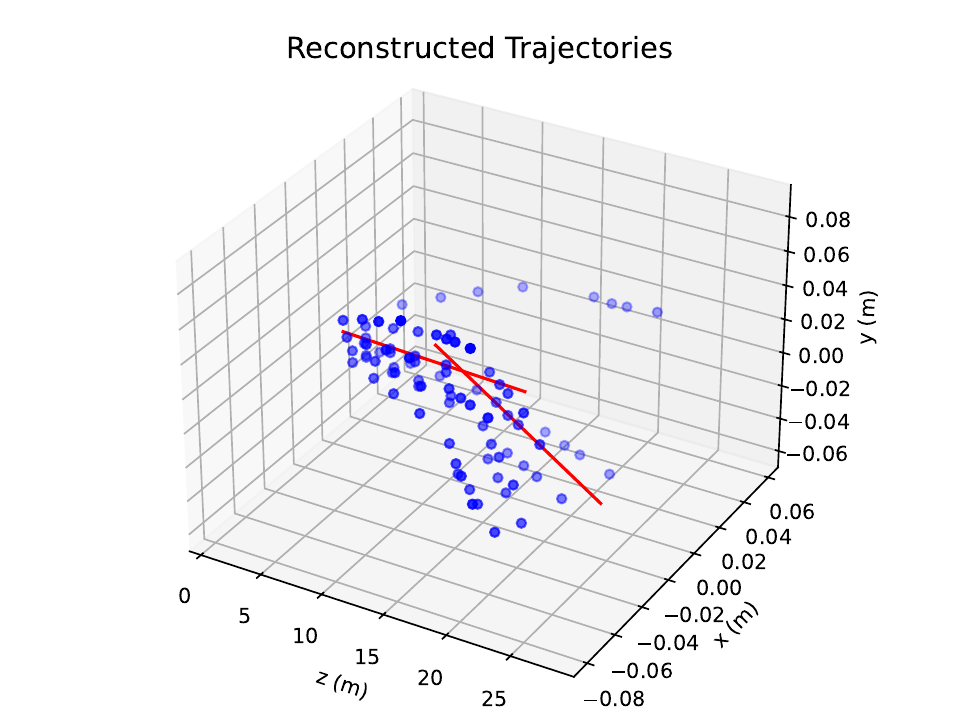}

    \caption{The top left figure shows the process of line fitting in Region A and right figure, Region C. Bottom figure shows the reconstructed trajectories in the two regions. The color of the line correspond to goodness of the fit or residuals and a darker color indicates lower residuals}

    \label{fig:spectro_line_fitting}
\end{figure}

Figure~\ref{fig:spectro_line_fitting} visualizes the process of line fitting. It is performed as follows:
\begin{enumerate}
    \item \textbf{Initial Setup}:
    \begin{itemize}
        \item Begin by selecting the hit in the first tracking station using the truth information.
        \item Perform a straight-line fit between this hit and all hits in the second tracking station, using the bin centers of the second station.
    \end{itemize}

    \item \textbf{Accounting for Uncertainty}:
    \begin{itemize}
        \item While fitting the line, the uncertainty associated with the bin width is incorporated into the calculations. The bin width uncertainty is accounted for by considering the standard deviation of a uniform distribution over the bin width in both the \(x\) and \(y\) dimensions, computed as \(\sqrt{\frac{2}{12}} \cdot \frac{1}{B_{i+1}}\), where \(B_{i+1}\) represents the granularity of the bins in the \(i\)-th station. This single factor accounts for contributions from both $x$ and $y$ dimensions. No uncertainty is assumed in the \(z\) dimension.
    \end{itemize}

    \item \textbf{Iterative Line Fitting}:
    \begin{itemize}
        \item Identify the closest hit in the next tracking station based on the initial fit.
        \item Perform a line fit that includes this newly identified hit, taking into account the uncertainty associated with the granularity of the new tracking station.
        \item Repeat this process iteratively until hits from all tracking stations in Region A have been included in the fit.
    \end{itemize}

    \item \textbf{Selecting the Best Fit}:
    \begin{itemize}
        \item Evaluate all possible line fits for Region A (considering the hits in the second tracking station) and select the one with the lowest residuals as the predicted trajectory.
    \end{itemize}

    \item \textbf{Region C Fit}:
    \begin{itemize}
        \item Repeat the above process for Region C. Start by taking the first hit from the truth information to avoid complexities such as the fringe field effects.
    \end{itemize}

    \item \textbf{Momentum Calculation in the \((x,z)\) Plane}:
    \begin{itemize}
        \item Using the trajectories determined for Region A and Region C, compute the momentum in the \((x,z)\) plane ($p_{XZ}$) using Equation~\ref{eqn:sin_eqn}.
    \end{itemize}

    \item \textbf{Momentum Calculation in the \(y\) Plane and Overall Momentum}:
    \begin{itemize}
        \item Use the trajectory to compute the momentum in the \(y\) plane ($p_Y$).
        \item Combine $p_{XZ}$ and $p_Y$ to calculate the particle's total momentum ($p$).
    \end{itemize}
\end{enumerate}

To achieve high performance, we utilized CUDA programming and PyTorch vectorization, enabling us to test a single design in approximately $0.06$~seconds, employing $0.12$M events and $1.32$M particles.


Figure~\ref{fig:results_spectro} shows the training process. The reward is shown as a function of the number of designs tested and it can be observed the agent quickly learns how the spectrometer looks like. At the start, the reward is very low ($<300$) as less than required number of tracking stations have been placed. Further, this results in constant values and discontinuous jumps in the reward function. This is not a problem for RL where the agents know how to successfully navigate such spaces. It can also be observed that the resolution and efficiency closely correspond to the reward. Here, the breaks at the start also correspond to the region where these metrics are undefined as enough tracking stations have not yet been placed.

The same figure also shows the best designs at different training stages. In the best designs, the tracking stations are not always placed to maximize the span. This is because if placed too early, the multiple scattering will worsen the resolution. Therefore, the agent has to learn to balance multiple factors and choose the best design.

Table~\ref{tab:spectro_results} show the results quantitatively. The results are also compared with two baseline design. The first baseline design, three equidistant tracking stations are placed in both Region A and Region C. In the second baseline design, four such tracking stations are placed. It can be observed that the spectrometer designed by the RL agent has significantly better performance.

\begin{table}[h!]
\centering
\begin{tabular}{|l|l|l|l|l|}
\hline
\textbf{} & \textbf{10 GeV Eff} & \textbf{10 GeV Res} & \textbf{100 GeV Eff} & \textbf{100 GeV Res} \\ \hline
\textbf{Baseline design 1} & $89.0695 \pm 0.1275$ & $6.48 \pm 0.03$ & $98.0982 \pm 0.0558$ & $13.83 \pm 0.06$ \\ \hline
\textbf{Baseline design 2} & $93.7342 \pm 0.0989$ & $5.97 \pm 0.03$ & $99.1317 \pm 0.0379$ & $13.15 \pm 0.05$ \\ \hline
\textbf{RL design} & $100.0000 \pm 0.0000$ & $3.74 \pm 0.02$ & $99.9417 \pm 0.0099$ & $7.87 \pm 0.03$ \\ \hline
\end{tabular}
\caption{Spectrometer resolution and efficiency of the RL based design vs two reference baseline designs. All numbers are in percentages}
\label{tab:spectro_results}
\end{table}

\section{Outlook}
The presented studies demonstrate the applicability of RL for instrument design using simple examples. In this section, we discuss how these works can be extended for more complex instrument design problems.

\subsection{Flexible Reconstruction Algorithms}
In the study presented on spectrometer design, a straightforward reconstruction algorithm was employed. However, for scenarios involving more generic and irregular detector geometries, such algorithms often prove insufficient. In such cases, advanced approaches like graph neural networks (GNNs) can provide a significant advantage. For instance, \textcite{qasim2022end} demonstrated the application of dynamic graph neural networks~\parencite{Qasim2019} for end-to-end reconstruction tasks by utilizing the Object Condensation technique~\parencite{Kieseler2020}. This method has shown robust performance in handling complex and irregular environments, making it a promising tool for detector design optimization.

One potential application of these advanced methods is in optimizing the transverse segmentation of calorimeters, where traditional algorithms may fall short. Additionally, as evidenced by the work of \textcite{dgnns_for_track}, dynamic graph neural networks can also be track reconstruction. These include scenarios such as designing tracker systems that must operate within magnetic fields, where the simpler reconstruction algorithms described in Section~\ref{sec:spectrometer_design_study} are not suitable.

Moreover, such graph-based algorithms offer the flexibility to be jointly optimized alongside RL agents’ policies. This simultaneous optimization could enable a more integrated approach to detector design, leveraging the strengths of both advanced machine learning techniques and physics-driven optimization frameworks. These developments highlight the potential of GNNs to revolutionize tasks that require intricate geometrical reasoning and adaptive design strategies in experimental physics.

\subsection{Amortization of Reward and Cost}
In the presented studies, a large number of simulations are performed to get a good estimate of the design score. This makes it convenient for the RL algorithm as the variance in the reward function is low. As an example, the SHiP Muon Shield aims to reduce the flux rate by over six orders of magnitude and if a design under consideration already gives multiple resulting hits after testing a few hundred,  this design can be easily ruled out. Similar argument can be made also for the calorimeter. Modern RL algorithms are designed to operate in the conditions of such high variance returns i.e. the reward function is stochastic. Employing this property in future work can award computational benefits making the discovery process quicker.

\subsection{Off-policy algorithms}
In this work, PPO has been used, which is an on-policy algorithm. This implies that any time the policy changes, the agent must collect a new set of experiences based on the updated policy to continue learning effectively and discard the old ones (called experience replay in RL terminology). Therefore, the algorithm is not the most sample efficient. This was as choice of convenience in the conducted study to quickly get results. However, techniques such as prioritized experience replay \cite{priotri_experience_replay} will allow the use of off-policy algorithms which are more computationally efficient. As an alternate, on-policy but continual learners can also be employed instead which don't suffer from catastrophic forgetting \cite{catastrophic}, a particularly pronounced problem faced by neural networks.

\subsection{Better Policy and Value Networks, and Improving Observability:} In the presented experiment of calorimeter design, the observation space was only two dimensional, representing the longitudinal dimension as well as the thickness budget used, which forces the agent to operate in a highly partially observable environment. Given that the full design is known, one can improve this by employing a better representation of the current state. For only longitudinal segmentation, this can be represented as a vector, as in the example of the spectrometer design but in end-to-end optimization, a more generic graph neural network could be used. For instance, a GNN was employed by \textcite{mirhoseini2021graph} for the chip placement problem. They will also provide more powerful policy and value network compared to the fully connected neural networks a choice of convenience in the experiments presented in this work.

\subsection{Surrogate simulators}
The computational expense of Monte Carlo simulators, particularly in propagating particles through dense materials, has spurred extensive research into faster machine learning-based simulation alternatives. Initially focused on simulating calorimeter showers~\parencite{paganini2018calogan} using generative adversarial networks (GANs)~\parencite{goodfellow2014generative}, recent advances have expanded into the fast simulation of particle jets~\parencite{touranakou2022particle} and other final-state observables. Researchers have explored a variety of generative models, including Variational Autoencoders (VAEs)~\parencite{Kingma2014}, GANs, Normalizing Flows~\parencite{rezende2015variational}, and other deep generative architectures, each contributing unique advantages in accuracy and efficiency. These ML-driven methods now play a vital role in reducing simulation time, making it feasible to handle the large-scale data demands of high-energy physics experiments while maintaining high fidelity in modeling complex particle interactions and responses within detectors.

Even more recently, research has explored ML-based fast simulation of detector agnostic responses~\parencite{buhmann2023caloclouds, buhmann2024caloclouds}. Here, diffusion models~\parencite{ho2020denoising} were employed to generate the detector responses as point clouds, and this can be extended to non-uniform sampling calorimeters where regular grid structures are absent. It is also possible to extend this work to condition it on detector configuration so one can learn the truly detector agnostic response.

\section{Conclusion}
\label{sec:conclusion}
In this work, we explored and demonstrated the application of Reinforcement Learning (RL) as a novel approach for instrument design, showcasing its effectiveness through two distinct empirical studies. The first study focused on the design of a calorimeter, while the second targeted the design of a spectrometer. Both studies utilized a RL agent that employed a mixed action space framework to construct these instruments layer by layer, simulating a practical, iterative design process. This mixed action space combined continuous and discrete actions to account for the diverse aspects of the design process, thereby allowing the agent to make comprehensive decisions.

In the context of the calorimeter and spectrometer designs, the continuous actions allowed the agent to determine optimal placements of the detector components (i.e.  active layers or tracking stations). These components are integral to the functionality of the instruments, influencing factors such as resolution and detection efficiency. On the other hand, the discrete actions enabled the agent to make categorical decisions about the physical properties of these components, such as the granularity of the sensor layers or the thickness of the detectors. By combining these two types of actions, the RL agent could effectively navigate the complex and multidimensional design space without requiring extensive prior knowledge or predefined templates.

We employed straightforward yet effective reward functions to guide the agent during the design process. These reward functions quantitatively assessed the performance of the proposed designs based on relevant metrics, such as detection accuracy, energy resolution, or overall efficiency. The simplicity of these reward functions underscores the robustness and adaptability of our methodology. Remarkably, the RL agent was able to autonomously generate high-performing designs for both instruments, showcasing its ability to learn and optimize effectively even in the absence of significant domain-specific priors.

This study highlights the  advantages of RL in addressing instrument design challenges, particularly when compared to differentiable optimization methods. Differentiable approaches typically require a well-defined, parameterized model of the instrument, which can limit their applicability in scenarios where such models are unavailable or difficult to construct. In contrast, our RL-based methodology circumvents these constraints by directly interacting with the design environment and placing sequentially. Further, through exploration and trial-and-error learning, the local minima are also effectively avoided. However, the differentiable approaches could still potentially complement the proposed method by fine-tuning the proposed design to make it final.

Looking ahead, we see tremendous opportunities to extend this approach to even more intricate design challenges. Future research will aim to apply this methodology to the development of more complex instruments, potentially involving multi-objective optimization and additional constraints that better reflect real-world scenarios. By integrating advanced RL techniques, such as hierarchical RL or model-based approaches, we aspire to further enhance the capabilities and versatility of this design paradigm. These advancements will add tremendous value to future projects like the Future Circular Collider (FCC), where achieving unprecedented energy scales will require the most optimized instrumentation.
\bibliographystyle{plainnat}
\bibliography{references}

\end{document}